\documentstyle[prl,aps,twoside,multicol,rotate,epsf]{revtex}

\begin{document}

\title{How to Teleport Superpositions of Chiral Amplitudes}
\author{Christopher S. Maierle, Daniel A. Lidar and Robert A. Harris}
\address{Department of Chemistry, The University of California, Berkeley,\\
CA 94720}
\date{\today}
\maketitle

\begin{abstract}
Chiral molecules may exist in superpositions of left- and right-handed
states. We show how the amplitudes of such superpositions may be
teleported to the polarization degrees of freedom of a photon and thus
measured. Two 
experimental schemes are proposed, one leading to perfect, the other
to state-dependent teleportation. Both methods yield complete
information about the amplitudes.
\end{abstract}

\begin{multicols}{2}
\markboth{{Maierle, Lidar and Harris}}{{Maierle, Lidar and Harris}}

\section{Introduction}

``Quantum teleportation,'' proposed in 1993 by Bennett et
al. \cite{Bennett:93}, has become a reality
\cite{Bouwmeester:97,Boschi:98}. However, to date all of the
teleportation work deals with {\em 
intra}-species teleportation (e.g., atom-atom or photon-photon). In
this Letter we propose an inter-species teleportation scheme.
Specifically, we outline experiments in which the information
contained in a superposition of chiral amplitudes: $|\phi _{M}\rangle
=a|L\rangle +b|R\rangle$,
may be teleported to a photon. Here, $|L\rangle $ and $|R\rangle $ are
the left and right handed states of a chiral molecule. In special
cases, the teleportation scheme presented here can also be used to 
teleport the state of more general molecular superpositions such as
superpositions of cis and trans isomers.
While methods for creating and detecting a molecular state of the form
$|\phi _{M}\rangle$
have already been discussed
\cite{Quack:86;Harris:94b;Cina:94;Cina:95,Harris:94}, the
corresponding experiments have not been performed. Indeed,
no chiral superposition has ever been measured. By teleporting the
information contained in the amplitudes $a$ and $b$ to the
polarization vector of a photon, the superposition becomes easy to
detect via standard photon polarization measurements. Then by
performing another 
teleportation to a spin 1/2 nucleus or a trapped ion (as
envisioned in \cite{Bouwmeester:97}) one can imprint the chirality
information onto a much more stable state, suitable for further
manipulations: the nucleus or ion acts as a {\em quantum memory}
device. 

To teleport the
chiral superposition state $|\phi _{M}\rangle$
we take our entangled
pair to be two photons in the state:
$|\Psi _{12}^{-}\rangle =\frac{1}{\sqrt{2}}(|l_{1}\rangle |r_{2}\rangle
-|r_{1}\rangle |l_{2}\rangle )$.
Here, $|l\rangle $ and $|r\rangle $ denote left and right circularly
polarized photons. Photon 1 will become
entangled with the molecule and photon 2 will be the one whose
polarization state will receive the amplitudes $a$ and $b$. Initially,
the total molecule-photons state is unentangled: $|\psi \rangle =|\phi
\rangle |\Psi _{12}^{-}\rangle$.
As in \cite{Bennett:93}, this state can be rewritten as:

\begin{equation}
|\psi \rangle =\frac{1}{2} \left( |\Psi _{M1}^{-}\rangle |1\rangle +|\Psi
_{M1}^{+}\rangle |2\rangle +|\Phi _{M1}^{-}\rangle |3\rangle +|\Phi
_{M1}^{+}\rangle |4\rangle \right).
\label{eq:fullstatenum}
\end{equation}
Here the four maximally entangled ``Bell states'' are:
$|\Psi _{M1}^{\pm }\rangle = \frac{1}{\sqrt{2}}(|L\rangle |r_{1}\rangle \pm
|R\rangle |l_{1}\rangle )$ , 
$|\Phi _{M1}^{\pm }\rangle = \frac{1}{\sqrt{2}}(|L\rangle |l_{1}\rangle \pm
|R\rangle |r_{1}\rangle )$ ,
and the four states involving photon 2 are:
\begin{eqnarray}
|1\rangle & = & -{a \choose b}=-a|\,l_{2}\rangle -b|\,r_{2}\rangle \\
|2\rangle & = & -\sigma _{z}{a \choose b} =-a|\,l_{2}\rangle
+b|\,r_{2}\rangle 
\label{eq:1+2}
\\
\,|3\rangle & = & \sigma _{x}
{a \choose b}=b|\,l_{2}\rangle +a|\,r_{2}\rangle \\
|4\rangle & = & -i\,\sigma _{y}
{a \choose b} =-b|\,l_{2}\rangle +a|\,r_{2}\rangle .
\label{eq:3+4}
\end{eqnarray}
The $\sigma $'s are the Pauli matrices and $|\,l_{2}\rangle = {1
\choose 0} $, $|\,r_{2}\rangle = {0 \choose 1}$. As emphasized in
\cite{Bennett:93}, the above form implies that the
``teleportee'' photon (2) can be transformed into the state ${a
\choose b}$ by one of four simple unitary operations. Which of the
four operations needs to be applied depends {\em only} on the
measurement outcome of the projection onto the Bell states. The scheme
thus requires two bits of classical communication to transfer the
information regarding the {\em continuum} of quantum states ${a
\choose b}$, at the price of establishing prior entanglement. However,
this is not to say that a unitary transformation of the original
teleported photon state is the only way to obtain complete
information. Below we give an example of
``state-dependent'' teleportation which nonetheless does yield
complete information.  Note further that the circular polarized basis
plays no special role in the above discussion. We now show how to
extend the previous work by proposing an 
apparatus which can be used to teleport the superposition of chiral
amplitudes.

{\it Teleportation of Chirality} ---.
Consider the apparatus shown in Figure 1. Before photon
1 reaches the interferometer, the system is in the direct product
state
$|\phi \rangle |\Psi _{12}^{-}\rangle$. At some later time, photon 1 will have
reached the beam-splitter and thus will have some amplitude to be
found in the top arm and some amplitude to be found in the bottom arm
where it will interact with the molecule. 

Now it is well known that left and right circularly polarized photons
acquire different phase shifts when scattering through a chiral
molecule \cite{Barron:82}.  Ordinarily the phase shifts due to a single
molecule are undetectably small.  However, it has been shown that
by utilizing a high-finesse optical resonator (cavity), phase shifts
due to the coupling of a single photon to a single cesium atom may be as
large as $16^\circ$ \cite{Turchette:95}. The interaction which gives
rise to this phase shift is an electric dipole-dipole scattering
process, whereas optical activity is mediated
by an electric dipole-magnetic dipole interaction. Since molecular
magnetic dipole moments are about $10^{-2}$ smaller than electric dipole
moments, we
expect that with current technology, the phase shift due to optical activity
in a high-finesse cavity would be on the order of a tenth of a degree,
perhaps too small to be useful in our scheme. On the other hand, the
last decade has seen tremendous progress in the fabrication of
high-finesse optical cavities\cite{Kimble:94}  and we expect
that our proposed experiment will be feasible in the future.

Furthermore, natural optical activity cannot be enhanced by a standing
wave cavity because it is erased upon reflection back
through the optically active medium \cite{Landau+Lifshitz}.  
Thus, natural optical
activity is enhanced only by use of a {\em ring}-cavity. Another
possibility is 
to utilize the effect of E-field optical activity \cite{Harris:94}.  
This type of optical
activity does not vanish upon reflection through the medium so a standing
wave cavity may be used.  The formalism which we present here is applicable
to natural optical activity.  For the case of E-field optical activity, the
superposition of chiral states should be written in the `false chirality'
basis made from the states $|L \rangle \pm i|R \rangle$ \cite{Barron:86}.
These states are converted into one another by inversion and they are
to E-field  
optical activity what the states $|L \rangle$ and $|R \rangle$ are to
natural optical activity \cite{Harris:94}.  

The photon-molecule scattering implements a {\em conditional
phase-shift} mechanism.
Specifically:

\begin{eqnarray}
|l\rangle |L\rangle & \rightarrow & e^{i(kz-\varphi )}|l\rangle
|L\rangle \text{ 
; \ \ \ \ }|r\rangle |R\rangle \rightarrow e^{i(kz-\varphi )}|r\rangle
|R\rangle  \label{eq:PS0} \\
|l\rangle |R\rangle & \rightarrow & e^{i(kz+\varphi )}|l\rangle
|R\rangle \text{ 
; \ \ \ \ }|r\rangle |L\rangle \rightarrow e^{i(kz+\varphi )}|r\rangle
|L\rangle .
\label{eq:PS}
\end{eqnarray}
Here $e^{\pm i\varphi }$ is the phase shift due to optical activity
and $z$ is the free-space optical path length. For arbitrary molecular
superpositions, the Faraday effect\cite{Barron:82} gives rise to a table 
similar to the one above. However, with the Faraday effect,
the two different molecular
states give rise to different rotations and spin-independent indices of
refraction (related to the quantities $\phi$ and $kz$ in the above equations).
For the special case where the two states give equal and opposite rotations 
of the polarization vector and also give rise to the same spin-independent
phase shift, the Faraday effect gives rise to a set of equations identical
to the ones above but where $|L \rangle$ and $|R \rangle$ are replaced by
kets representing the two molecular states which are superposed.  For this
special situation, the Faraday effect can be used in place of optical activity
and the teleportation scheme can be used to teleport more general types of
molecular superpositions.  As an added bonus, the Faraday effect does not
vanish upon reflecting the beam back through the medium and can therefore be enhanced in a standing wave cavity.

In any case, using the above equations, it is easy to show that after the 
interaction with
the chiral superposition, the amplitude on the bottom arm of the
interferometer (Figure 1) is proportional to:

\begin{eqnarray}
|\psi _{\text{bot}}\rangle &\propto& \frac{e^{i\varphi }}{2}\left[ |\Psi
_{M1}^{-}\rangle |1\rangle +|\Psi _{M1}^{+}\rangle |2\rangle \right]
\nonumber \\
&+& \frac{
e^{-i\varphi }}{2}\left[ |\Phi _{M1}^{-}\rangle |3\rangle +|\Phi
_{M1}^{+}\rangle |4\rangle \right]
\label{eq:psi-bot}
\end{eqnarray}
The amplitude for going through the upper arm is still described by
Eq.~(\ref {eq:fullstatenum}) and the full state is a superposition 
of the
amplitude on the top and bottom arms. Now, by adjusting the path-length and
thus the phase of the amplitude in the top arm of the interferometer,
we can arrange so that only one pair of the states $|\Psi _{M1}^{\pm
}\rangle $ and $|\Phi _{M1}^{\pm }\rangle $ has non-zero amplitude to
reach the detector 2. Suppose we adjust the top arm so that only the
$|\Psi _{M1}^{\pm }\rangle $ reach detector 2. Then after the photon
has left the interferometer the state can be written (up to an overall
phase):
$|\psi \rangle =\frac{\sin (\varphi )}{\sqrt{2}}\left[ |\Psi
_{M1}^{-}\rangle _{ 
\text{D}_{2}}|1\rangle +|\Psi _{M1}^{+}\rangle _{\text{D}_{2}}|2\rangle 
\right] +|\psi' \rangle$.

Here the subscript D$_{2}$ indicates that the photon in that state is
heading for detector 2. The state $| \psi' \rangle $ is the amplitude
which in Figure 1 is now traveling vertically away from this detector
(denoted ``other''). We
are not concerned with the precise form of $| \psi' \rangle $; for
our purposes it suffices to know that this state involves photon amplitude
which will never intersect D$_{2}$. Using the basis of parity eigenstates $
|\pm \rangle $ and linear polarizations $|x\rangle $, $|y\rangle $: 

\begin{eqnarray}
|\pm \rangle &=&\frac{1}{\sqrt{2}}(|L\rangle \pm |R\rangle ) \\
|x\rangle &=&\frac{1}{\sqrt{2}}(|l\rangle +|r\rangle ) \:;\:\:
|y\rangle =\frac{i}{\sqrt{2}}(|l\rangle -|r\rangle )\text{ ,}
\end{eqnarray}
we can rewrite the post-interferometer state as: 

\begin{eqnarray}
|\psi \rangle & = & \frac{\sin (\varphi
)}{\sqrt{2}}(\frac{1}{\sqrt{2}}|+\rangle 
|x\rangle _{\text{D}_{2}}-\frac{i}{\sqrt{2}}|-\rangle |y\rangle _{\text{D}
_{2}})|1\rangle   \nonumber \\
& + &(\frac{1}{\sqrt{2}}|-\rangle |x\rangle _{\text{D}_{2}}-\frac{i}{\sqrt{2}}
|+\rangle |y\rangle _{\text{D}_{2}})|2\rangle +|\psi' \rangle \text{ .}
\end{eqnarray}
The laser in Figure 1 is used to produce a $\pi$ pulse, tuned to a
transition between the ground state of the molecule and an excited state of
definite parity which fluoresces. Suppose that the excited state is of odd
parity; in the electric dipole approximation the parity must change
upon electronic excitation, so only the state $|+\rangle $ will be excited.
Therefore, after excitation and fluorescence, we arrive at the state: 

\begin{eqnarray}
|\psi \rangle & = & \frac{\sin (\varphi )}{\sqrt{2}} \left[
\left( \frac{1}{\sqrt{2}}|+\rangle 
|\nu _{1}\rangle |x\rangle _{\text{D}_{2}}-\frac{i}{\sqrt{2}}|-\rangle |\nu
_{0}\rangle |y\rangle _{\text{D}_{2}} \right) |1\rangle \right.  \nonumber \\
& + & \left. \left( \frac{1}{\sqrt{2}}|-\rangle |\nu _{0}\rangle |x\rangle _{\text{D}_{2}}-
\frac{i}{\sqrt{2}}|+\rangle |\nu _{1}\rangle |y\rangle _{\text{D}
_{2}} \right) |2\rangle \right] +|\psi' \rangle . \nonumber \\
\end{eqnarray}
The $|+\rangle $ molecular state has become coupled to a spontaneously
emitted photon ($|\nu_{1}\rangle $), whereas the $|-\rangle $ state
has not (the vacuum state $|\nu _{0}\rangle $). Teleportation can now
be performed by projecting onto an {\em unentangled} state
\cite{Cirac:94}: one places a polaroid oriented in the $x$ 
direction in front of the detector 2 and looks for coincidences with
detector 1 (which detects the spontaneous emission $|\nu
_{1}\rangle$). A coincidence measurement then constitutes a projection
onto the state $|\nu _{1}\rangle |x\rangle _{\text{D}_{2}}$. This
implies that the 
teleportee photon is in the state $|1\rangle =-a|l_{2}\rangle
-b|r_{2}\rangle $. Teleportation has been achieved. By altering the length
of the top arm of the interferometer and/or exciting the molecule to an even
parity state, we can teleport to any of the four photon states in
Eq.~(\ref{eq:fullstatenum}).

{\it State-Dependent Teleportation} ---.
The latter scheme accomplishes perfect
teleportation: an {\em unknown} amplitude of the chiral superposition
appears in the polarization vector of photon 2. We
will next consider a simplified experiment which avoids the use of
interferometry, and accomplishes {\em state-dependent}, imperfect
teleportation.

Suppose we remove the upper arm of the interferometer. Then the entire
apparatus is represented by $|\psi _{\text{bot}}\rangle $ as in Eq.~(\ref
{eq:psi-bot}). We again rewrite the amplitude in the $|+\rangle $, $
|-\rangle $, $|x\rangle $ and $|y\rangle $ representation, yielding:
$|\psi _{\text{bot}}^{\prime }\rangle =\left[ |-\rangle |y\rangle |1^{\prime
}\rangle +|+\rangle |y\rangle |2^{\prime }\rangle +|-\rangle |x\rangle
|3^{\prime }\rangle +|+\rangle |x\rangle |4^{\prime }\rangle \right]$
where the four unnormalized teleportee photon states are:

\begin{eqnarray}
|1^{\prime }\rangle  &\equiv &\frac{i}{2\sqrt{2}}\left( \,e^{-i\varphi
}|2\rangle -\,e^{i\varphi }|4\rangle \right) 
\label{eq:1'} \\
|2^{\prime }\rangle  &\equiv &\frac{i}{2\sqrt{2}}\left( \,e^{-i\varphi
}|1\rangle -\,e^{i\varphi }|3\rangle \right) 
\label{eq:2'} \\
|3^{\prime }\rangle  &\equiv &\frac{1}{2\sqrt{2}}\left( \,e^{-i\varphi
}|1\rangle +\,e^{i\varphi }|3\rangle \right) 
\label{eq:3'} \\
|4^{\prime }\rangle  &\equiv &\frac{1}{2\sqrt{2}}\left( \,e^{-i\varphi
}|2\rangle +\,e^{i\varphi }|4\rangle \right)
\label{eq:4'}
\end{eqnarray}

Denoting $a=\alpha e^{i\theta _{a}}$ and $b=\beta
e^{i\theta _{b}}$, the norms are:
\begin{eqnarray}
\text{Pr}(1^{\prime }) &=&|\langle \psi _{\text{bot}}^{\prime }|-,y\rangle
|^{2}=\text{Pr}(3^{\prime })=|\langle \psi _{\text{bot}}^{\prime
}|-,x\rangle |^{2} \nonumber \\
&=& \frac{1}{4}[1-2\alpha \beta \cos \left( \theta
_{a}-\theta _{b})\cos (2\varphi \right) ] \label{eq:norm1} \\
\text{Pr}(2^{\prime }) &=&|\langle \psi _{\text{bot}}^{\prime }|+,y\rangle
|^{2}=\text{Pr}(4^{\prime })=|\langle \psi _{\text{bot}}^{\prime
}|+,x\rangle |^{2} \nonumber \\
&=& \frac{1}{4}[1+2\alpha \beta \cos \left( \theta
_{a}-\theta _{b})\cos (2\varphi \right) ]\text{.} \label{eq:norm2} 
\end{eqnarray}
We notice that the transformation matrices that send ${a \choose b}$
into $|1'\rangle$ through $|4'\rangle$ are not unitary. For example,

\[
|1^{\prime }\rangle  
= \frac{1}{i 2\sqrt{2}}\left( 
\begin{array}{cc}
e^{-i\varphi } & -e^{i\varphi } \\ 
e^{i\varphi } & -e^{-i\varphi }
\end{array}
\right) \left( 
\begin{array}{c}
a \\ 
b
\end{array}
\right).
\]
The resulting
states are, however, {\em pure}. Appropriately renormalized, they may
be represented by polarization vectors on the unit Bloch
sphere. Hence there is a unitary rotation which carries each
teleported vector into the original state, ${a \choose b}$. However,
as seen from Eqs. (\ref{eq:norm1}) and (\ref{eq:norm2}), the
transformation depends upon the values of $a,b$, and the phase-shift
angle $\varphi$.  In this sense the present scheme constitutes a
state-dependent, imperfect teleportation, since it cannot be used to
teleport an unknown quantum state \cite{Mor:96}.
Nevertheless, it is possible to obtain {\em full}
information about $a$ and $b$ by standard optical methods. One can measure
the relative phase and relative magnitude of the polarization components of
photon 2. The relative phases and separately, the relative magnitudes,
are equal in pairs. Thus
by transmission of a single bit of classical information, it is possible to
tell cases $1^{\prime },3^{\prime }$ from $2^{\prime },4^{\prime }$ and to
obtain complete information on the chiral superposition. Of course,
each measurement which contributes to this process destroys the
superposition of polarizations of photon 2. This loss cannot be prevented in
the perfect teleportation scheme either if the actual values of $a$
and $b$ are needed. From this perspective there is no real advantage
to the perfect scheme. Indeed, the perfect scheme is better only if photon 2
is put to use in a later quantum information processing stage, such as
an input to a quantum computer.

{\it Discussion} ---.
This work has, for the first time, considered in detail the possibility
of inter-species teleportation. This led us to propose a concrete
scheme by which the quantum chiral state of a single molecule could be
measured, by means of transferring the chiral information to an easily
measurable photon polarization state. We have outlined two
experiments, one leading to perfect, 
or unitary, teleportation of the amplitudes of a chiral superposition,
the other to state-dependent teleportation. However we have shown that
the latter is able to transmit, in general, full amplitude information
as well. The key to the schemes proposed here is the use of the
parity-conserving {\em symmetry} governing the interaction between
light and a chiral molecule, Eqs.~(\ref{eq:PS0}) and
(\ref{eq:PS}). This symmetry leads to the possibility of
implementing a conditional phase-shift, without which teleportation
cannot take place. We conjecture that it is possible to exploit other
symmetries in order to affect inter-species teleportation in other
cases. Indeed, we have discussed
how, using the Faraday effect, the state of a more general molecular
superposition 
can be teleported. In the same vein, any other spin $1/2$ particle can
be used to replace the photons in our scheme,
but with a different interaction, such as spin-orbit coupling
\cite{Maierle:97}.

A virtue of the
method of chiral teleportation is that it provides a genuine new way
of measuring chiral superpositions of chiral amplitudes. Indeed, even
if the original molecular state is $|L \rangle$, successful
teleportation is a manifestation of at least one pair of
superpositions, $|+\rangle$ and $|-\rangle$.

Model calculations show that chiral
superpositions in media 
are extremely short-lived: they decohere on a time-scale of pico- to
femtoseconds \cite{Cina:94a}. Hence one might wonder whether the
chiral superposition will not decohere over the time scale needed for the
photon to interact with it. Collisions with the walls and asymmetries
of the cavity are the chief decoherence agents as they lead to
fluctuating chiral environments \cite{Harris:81}. In the case of a
high-finesse cavity we estimate that this should lead to decoherence
times of the order of 1 second. Decoherence thus does
not present a significant obstacle in accomplishing the proposed experiment.

Another issue brought up by this work is the possibility of
probing quantum properties in ``large'' objects, and thus the
transition (be it continuous or sharp) as a function of object size to
classical behavior. The emergence of the latter is one of the most
fascinating unsolved problems of present-day physics. At which point
is the object ``too large'' to enable teleportation of its chiral
superposition? The number of degrees of freedom of the object will set
the decoherence time-scale, since it determines the coupling to the
bath degrees of freedom. It thus controls the extent of
``environmental-monitoring,'' leading to classical behavior, i.e., the
absence of quantum interference in large objects
\cite{Zurek}. In principle, as long as the object
is chiral and there exist excited non-degenerate states of definite
parity, our schemes apply. {\it Ceteris paribus}, failure to teleport
may thus be taken as an indication of classicality of the chiral
object.

Finally, the extension of quantum teleportation to
superpositions of molecular states has yielded an entirely new way of
measuring given superpositions of chiral amplitudes.  It is therefore
tempting to consider {\em reverse} inter-species
teleportation. Consider, e.g., diasterioisomers: these are two
molecules connected by a chemical bond or a Van der Waals complex. The
molecules are are mirror images of one another. That is, one is
$|L\rangle$ and the other is $|R \rangle$. For example, a left-handed
oligomer of diphenyl alanine (DA) connected by a disulphide bridge
(DB) to a right-handed oligomer of DA. The DB bridge is easily
cleaved. Let us imagine that the dimer is cooled down to a $J=0$ state
of total angular momentum. When cleaved the dimer state breaks up into
monomer states of equal total angular momentum and equal and opposite
$z$ component of angular momentum, $M$.  The fragments also move in
opposite directions.  Clearly then there is a sum of entangled states
in $M$.  Using the same scheme as described above for a simultaneous
measurement on a chiral molecule and a photon, we can now teleport a
superposition of {\em photon} polarization states to create a
superposition of handed states in one of the molecules. Thus we can
create arbitrary superpositions of chiral
amplitudes through inter-species teleportation. 

\section*{Acknowledgements}
We would like to acknowledge interesting conversations with T. Lynn, 
and Profs. I. Tinoco, K.B. Whaley and W.H. Miller. This
work was supported by a grant to R.A.H. from the N.S.F.

\vspace{-1.5em}
\begin{figure}
\hspace{6em}
\vspace{2em}
\epsfysize=3cm
\epsfxsize=5.5cm
\epsffile{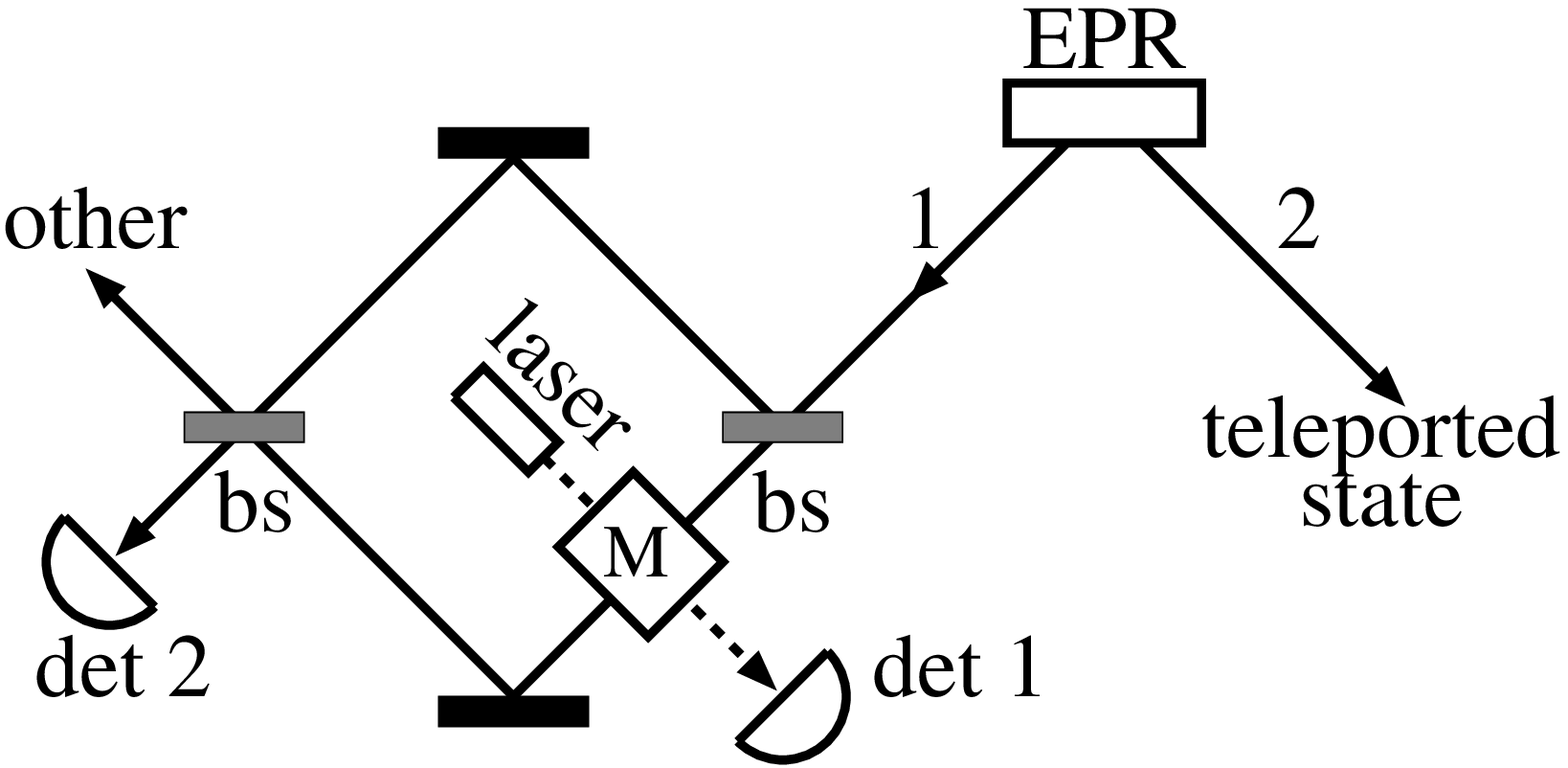}
\label{fig:expt}
\end{figure}

\vspace{-3.em}
\noindent Fig. 1:
{\small A source produces an entangled pair of photons, 1 and 2. Photon 1 
enters an interferometer and interacts with the molecule, labeled M.
A $\pi$ pulse, produced by the box labeled laser, excites the molecule
to a state of definite parity and the resulting fluorescence is
detected by detector 1. Detector 2 is set to detect photons of a
definite polarization. A coincidence measurement then teleports the
state of the molecular superposition to the polarization state of
photon 2.}

\end{multicols}

\end{document}